\documentclass[arxiv,floatfix,twocolumn,showpacs]{revtex4}
\usepackage{color} 
\usepackage{graphicx} 
\usepackage{amsmath, amsthm, amssymb}
\usepackage{amsthm}
\usepackage{multirow} 
\begin{document} 
\title{Ion Hydration and Associated Defects in Hydrogen Bond Network of Water: Observation of Reorientationally Slow Water Molecules Beyond First Hydration Shell in Aqueous Solutions of MgCl$_2$}

\author{Upayan Baul} 
\email{upayanb@imsc.res.in} 
\author{Satyavani Vemparala} 
\email{vani@imsc.res.in} 
\affiliation{The Institute of Mathematical Sciences, C.I.T. Campus, Taramani, Chennai 600113, India} 
\date{\today}

\begin{abstract} Effects of presence of ions, at moderate to high concentrations, on dynamical 
properties of water molecules are investigated through classical molecular dynamics simulations 
using two well known non-polarizable water models. Simulations 
reveal that the presence of magnesium chloride (MgCl$_2$) induces perturbations in the hydrogen bond 
network of water leading to the formation of bulk-like domains with \textquoteleft defect 
sites\textquoteright~on boundaries of such domains: water molecules at such defect sites have less 
number of hydrogen bonds than those in bulk water. Reorientational 
autocorrelation functions for dipole vectors of such defect water molecules are computed at different 
concentrations of ions and compared with system of pure water. Earlier experimental and simulation studies indicate significant differences 
in reorientational dynamics for water molecules in the first hydration shell of many dissolved ions. 
Results of this study suggest that defect water molecules, which are beyond the first hydration 
shells of ions, also experience significant slowing down of reorientation times as a function of 
concentration in the case of MgCl$_2$. However, addition of cesium chloride(CsCl) to water does not 
perturb the hydrogen bond network of water significantly even at higher concentrations. This 
difference in behavior between MgCl$_2$ and CsCl is consistent with the well-known Hofmeister series. 
\end{abstract}
\maketitle

\section{Introduction} Effects of simple inorganic salts on the molecular properties of water 
are at the heart of a vast number of interesting and complex processes such as the stability of 
proteins and nucleic acids~\cite{nostro} and environmentally relevant processes~\cite{buszek,jung}.  
Understanding the effect of dissolved ions on the structural and dynamical properties of water is 
essential in this regard. The Hofmeister effect, which includes highly ion-specific effects on 
aggregation dynamics of proteins~\cite{hof,kunz} and other biologically relevant 
processes~\cite{kunz,marcus,tobi} has been of significant 
interest lately. Spectroscopic techniques
have been instrumental in probing the cooperative ion hydration mechanism and consequent long-range 
structural and dynamical effects of certain salts, or ion combinations, on water~\cite{tiel,tiel2,obrien,obrien2,obrien3,vanpost,paschek}. Earlier experiments 
suggested that the effect of ions on dynamical properties of water is largely restricted to their 
first hydration shell~\cite{omta2003,omtabak}. However, recent experiments, using a combination of 
femtosecond time resolved infrared (fs-IR) and dielectric relaxation spectroscopy, have shown the 
existence of a fraction of reorientationally slow water molecules~\cite{tiel,tiel2} well beyond the first 
hydration shells of dissolved MgSO$_4$ ions.  In these experiments, two sub-populations of water molecules 
were identified for various salts: one with reorientation timescales comparable to pure water 
($\sim$2.6 ps) and the other showing characteristically slower reorientations ($\sim$10 ps). The 
fraction of total water molecules contributing to the sub-population of slow-water molecules was seen to 
increase with increase in salt concentration for all salts and the magnitude of increase was 
observed to be highly dependent on ion combinations: being largest for combinations of strongly 
hydrated ion species (Mg$^{2+}$, SO$_4^{2-}$).

Despite the recent experimental results~\cite{tiel}, the existence of long-range temporal effects of salts, 
beyond first hydration shell, is controversial owing to results from other 
experimental~\cite{omta2003,funkner2011,omtabak,giammanco} as well as 
simulation~\cite{verde,stirn,yanggao,linwater,funkner2011} studies which suggest the contrary. While 
an intense cooperative slowdown of water reorientation has been observed in presence of ions with 
high charge densities, the range of presence of slow water molecules has been found to be confined to the 
first hydration shell of the ions~\cite{verde,stirn,yanggao,linwater,funkner2011,omtabak}. The focus 
of present work is on the extended hydrogen bond network in bulk water and the domain formation and 
existence of \textquoteleft defect water molecules\textquoteright~at boundaries of such domains, 
when salt is added. Instead of classifying water molecules based on radially varying spherical 
hydration shells around ions, sub-populations depending on whether they are bulk-like or defect 
water molecules are considered. Using molecular dynamics (MD) simulations, we present evidence for 
reorientational slowdown of water molecules well beyond the first hydration shell for solutions of MgCl$_2$ 
and effect of salt concentration on the same. However for CsCl-water solutions, no significant 
effect on reorientational dynamics of water (irrespective of CsCl concentration) was found, which is
 in good agreement with the Hofmeister series.

\section{Methods} 
\subsection{System Setup}~Classical atomistic molecular dynamics (MD) simulations were performed 
with TIP4P-Ew~\cite{t4pew} and TIP3P~\cite{jorg} water models and using simulation package NAMD 2.9~\cite{namd}.
For simulations involving TIP4P-Ew water model, recently developed parameters for divalent ion Mg$^{2+}$ and optimized 
for correct coordination number~\cite{t4pew-di} were used. Monovalent (Cl$^-$, Cs$^+$) ion parameters for the 
TIP4P-Ew simulations were taken from extensively used halide and alkali ion parameters~\cite{t4pew-mono}. 
Standard CHARMM parameters were used for all ions~\cite{charmm} while simulating with TIP3P water. All systems 
were first equilibrated under constant pressure and temperature (NPT) and further under constant 
volume and energy (NVE) conditions. Production runs for all systems were performed under NVE conditions. For the 
NPT simulations, pressure was maintained at 1 atm using Langevin Piston~\cite{feller} and temperature coupling to 
external heat bath was used to maintain temperature at 298 K for simulations with TIP4P-Ew model and at 305 K 
for TIP3P. Lennard-Jones interactions were smoothly truncated with cutoff $12\,\mathring{A}$ using switching function 
between $10\,\mathring{A}$ and $12\,\mathring{A}$. Long-range electrostatic interactions were computed using particle 
mesh Ewald (PME) method. Timesteps of 1 and 2 fs were used for simulations involving TIP4P-Ew and TIP3P models respectively.\\
\emph{TIP4P-Ew water systems :} Initially a $54\mathring{A}\times54\mathring{A}\times54\mathring{A}$ box of water containing 5251 
water molecules was equilibrated for 5 ns under NPT conditions and further for 4 ns under NVE. \textit{Solvate} plugin of 
VMD~\cite{vmd} was used to produce MgCl$_2$ (2M, 3M, 4M) and CsCl (3M, 4M) solutions from this equilibrated configuration and each of 
the salt-water systems was equilibrated for 5 ns under NPT conditions and further 4 ns under NVE resulting in total equilibration 
time of 9 ns for each of six systems involving TIP4P-Ew water.\\
\emph{TIP3P water systems :} Initially a $50\mathring{A}\times50\mathring{A}\times50\mathring{A}$ box of water containing 4972 water 
molecules was equilibrated for 10 ns under NPT conditions. \textit{Solvate} plugin of VMD was used to produce 
salt solutions of the same concentrations as for TIP4P-Ew water molecules from this equilibrated configuration and each of 
these salt-water systems was equilibrated for 15 ns under NPT conditions. The pure water system was further simulated for 5 
ns, resulting in 15 ns NPT equilibration simulations for each of the six systems considered. All above systems were equilibrated 
for further 9 ns under NVE conditions resulting in total equilibration time of 24 ns for systems involving TIP3P water.\\
Further details of system setup are included in the TABLE SI of Supplementary Information (\textit{Supp.~Info.})~\cite{supplement}. 
Relatively high salt concentrations of salt were chosen in this study based on recently studied experimental 
concentrations~\cite{tiel2,giammanco} as well as to account for good statistics. For computation of 
reorientational autocorrelation functions and other dynamic observables, production runs of 1 ns 
each were carried out for all systems under NVE conditions, with system configurations saved every 0.1 ps. All the dynamical
analyses were performed over the last 0.5ns of the production run data. 

\subsection{Water domain identification}
\begin{figure}[ht]
 \includegraphics[scale=0.65]{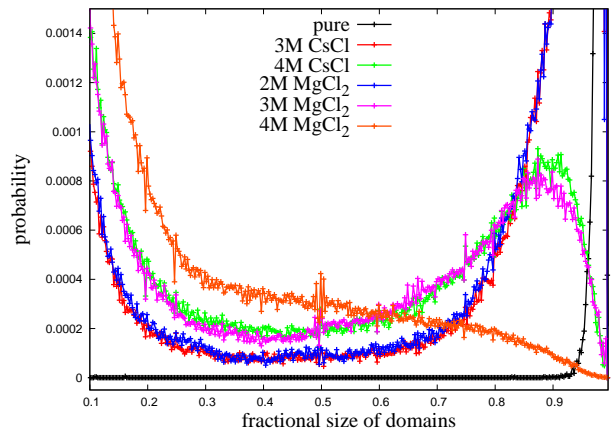}
 \caption{\footnotesize Domain size distribution plots scaled to $[0,1]$ for all systems with TIP4P-Ew water model
considered in the study and for slab thickness of 4$\mathring{A}$. The data are averaged over 5000 frames and 0.5 ns of simulation time.} 
\end{figure} 
To identify hydrogen bond (O-O distance $\leq$ 3.8$\mathring{A}$ and H-O${\cdot \cdot}$H angle $\leq$ 60$^{\circ}$ definition of hydrogen bond was used in the present study) network domains in all 
the twelve systems considered in this study, each system was divided into multiple overlapping slabs 
along each of the three orthonormal directions. Over each such slab, hydrogen bonded domains of 
water molecules were identified. The criterion for inclusion of water molecules to a single domain is
 the existence of a bidirected path between every pair, via a network of hydrogen bonds formed 
by water molecules in the same slab. Using this definition of water domain, it was observed that for 
pure water all values of slab thickness $\geq$2.7$\mathring{A}$ consistently resulted in a single 
spanning domain along the slab. Addition of salt to the water systems resulted in formation of 
multiple water domains, even when higher values of slab thickness (4$\mathring{A}$) were considered. 
The size distribution of different domains of TIP4P-Ew water systems for a slab thickness of (4$\mathring{A}$) and 
various concentrations of added salts, in comparison with the case of pure water is shown in FIG. 1. 
From the figure, it can be seen that for the case of pure water, a single spanning domain exists. 
With the addition of MgCl$_2$ salt, the size distribution of water domains depends on the 
concentration. For 2M salt concentration, the domain size distribution is similar to the pure water 
case, with additional appearance of smaller sized water domains. When the concentration of MgCl$_2$ 
is increased to 3M and above, the domain size distribution differs significantly from that of pure water. No 
single spanning domain remains in the system and the system primarily consists of many domains of 
water molecules of varying size distribution.Similar behavior is aslo observed when TIP3P water model is used and the 
results are given in \textit{Supp.~Info.} (FIG. SI)~\cite{supplement}. Ion specific local structural effects on water has been
widely studied in the literature~\cite{cappa,anders,omtabak,kally,soper,kally2}, generally
classifying ions into structure-makers and structure-breakers~\cite{marcus,marcus2,bakrev,kally2}
based on their effect on the hydrogen bond network of water. High charge density ions have been
known to exert strong patterning effects on first solvation shell water molecules leading to a
reduction in water-water hydrogen bonding~\cite{elcock,barbara,shinto}.

The domain size distribution in CsCl-water system, in contrast to MgCl$_2$-water systems,  is very similar to the pure water system even at 
4M concentration consistent with a recent ab initio MD study of other low-charge density salts such as NaCl and 
CsI~\cite{ding}. Following the identification of domains, 
water molecules residing at the domain boundaries were identified and recorded for all the three 
orthonormal directions. From this data, the water molecules which appear in all the three lists, 
corresponding to three orthonormal directions, are identified and labeled defect water molecules. This set of 
defect water molecules however may contain water molecules which are within the first hydration 
shell of any ions. From this super set of defect water molecules, a subset of water molecules which 
are not within the first hydration shell of any ion were identified, which are referred to 
henceforth as \textit{waterD} (pure water is referred to as \textit{waterP}). A further sub-population of defect water molecules were also identified: 
water molecules which are at domain boundaries in any two orthonormal directions (instead of 
three) and not in the first hydration shell of ions. These will be referred as \textit{waterD2} and constitute a more relaxed definition of defect water molecules and their sub-population
size would be larger than that of \textit{waterD} at moderate salt concentrations.

The probability distribution of local tetrahedral order parameter ($\textnormal{Q}$) was computed for \textit{waterP} and sub-populations \textit{waterD} 
for systems involving TIP4P-Ew model of water. $\textnormal{Q}$ for a water molecule is defined as~\cite{tethed1}

\begin{equation}
 \textnormal{Q}\,=\,\textnormal{1}\,-\,\frac{\textnormal{3}}{\textnormal{8}} \sum \limits_{\textnormal{j}=\textnormal{1}}^{\textnormal{3}} \sum \limits_{\textnormal{k}=\textnormal{j}+\textnormal{1}}^{\textnormal{4}} \left(\textnormal{cos} \varPsi_{\textnormal{jk}} \,+\, \frac{\textnormal{1}}{\textnormal{3}} \right)^\textnormal{2}
\end{equation}
where $\varPsi_{\textnormal{jk}}$ is the angle formed by the lines joining the oxygen atom of a given molecule and those of its nearest neighbours j and k 
$(\leq \textnormal{4})$. $\textnormal{Q}$ takes values 1 and 0 for ideal tetrahedral structures and ideal gas respectively. 

\section{Results}
\begin{figure}[h]
 \includegraphics[scale=0.6]{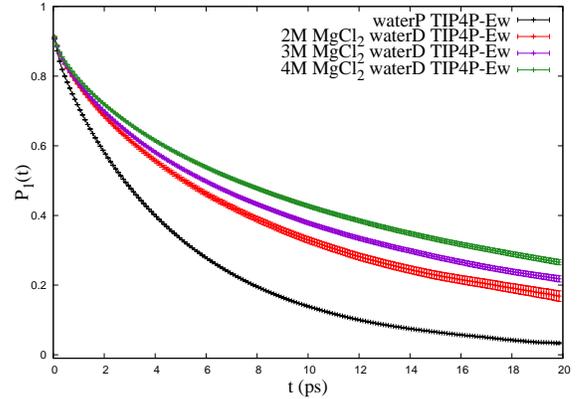}
 \caption{{\footnotesize Plots for P$_{\textnormal{1}}$(t) for \textit{waterD} in the presence of MgCl$_2$ (colored curves) at the concentrations studied and pure water (black) 
using TIP4P-Ew model. The error bars in all plots have been magnified 5 times}}
\end{figure}

The tetrahedrality results, in conjunction with hydrogen bonds per water molecules, suggest that the sub-population of \textit{waterD} molecules differ significantly from 
pure water molecules. The typical bimodal distributions~\cite{tethed1,tethed2} of tetrahedrality of water molecules, measured as in Eq(1), exhibit deviations in the case of defect water molecules (results
 are included in the \textit{Supp.~Info.} (FIG. SV)~\cite{supplement}. The number of hydrogen bonds per defect water molecule (\textit{waterD}) were found to be less than 2.
Reorientation autocorrelation functions P$_{\textnormal{1}}$(t) and P$_{\textnormal{2}}$(t) defined as first and second Legendre polynomials of 
water dipole vector ($\vec{p}$), a unit bisector of the H-O-H angle, were computed for various sub-populations of water molecules over 20 ps: 
\begin{eqnarray}
  \textnormal{P}_{\textnormal{1}}(\textnormal{t})\,&=&\,\langle \overrightarrow{p(\textnormal{t})}\ldotp \overrightarrow{p(0)} \rangle \\
  \textnormal{P}_{\textnormal{2}}(\textnormal{t})\,&=&\,\langle \frac{1}{2}(3 {\textnormal{cos}^2}(\overrightarrow{p(\textnormal{t})}\ldotp \overrightarrow{p(0)})-1) \rangle 
\end{eqnarray}
where the angular brackets denote average over number of water molecules in each sub-population and time. The errors for P$_{\textnormal{1}}$(t) and P$_{\textnormal{2}}$(t) have been 
obtained by computing standard deviations using block averages.\\
The computed correlation functions P$_{\textnormal{1}}$(t) for pure water and MgCl$_2$ solutions using TIP4P-Ew water model are plotted in FIG. 2 
(plots of P$_{\textnormal{2}}$(t) for TIP4P-Ew and both P$_{\textnormal{1}}$(t) and P$_{\textnormal{2}}$(t) for TIP3P are given in FIG. SII and FIG. SIII respectively in 
\textit{Supp.~Info.})~\cite{supplement}. A significant slowing down of reorientational times for sub-population \textit{waterD} as a function of salt concentration can be
seen from the figure. These results suggest that \textquoteleft slow water molecules\textquoteright~exist beyond the first hydration shell of both cations and anions. 
Earlier experiments and simulations show that the propensity of formation of ion-water clusters is higher at higher concentration of salts in water~\cite{allo}.
Water molecules trapped in such clusters can experience very slow reorientational times. It is to be noted that though the salt concentrations considered in this 
study are high, the sampling of sub-population \textit{waterD} will not include such trapped water molecules, 
by definition. The \textit{waterD} molecules considered can, at best, be part of three solvent separated ion pairs. Extensive studies of the reorientation of water 
molecules~\cite{laage,laage06,laage08,laagerev,verde12} have shown that autocorrelation functions of a body-set vector, including dipole vector, of water 
molecules involve distinct time scales. A fast, sub-picosecond decay due to librational motion followed by a slower component that can be attributed to structural 
changes such as hydrogen bond exchange and reorientations of hydrogen bonded water molecules. Multiexponential functions thus serve as good fit functions
for such correlations and they have been fit to either bi- ($A_3=0$) or tri-exponential functions of the form
P$_{\textnormal{1/2}}(t)\,=\,A_{1}\textnormal{exp}(-\frac{t}{\tau_{1}})\,+\,A_{2}\textnormal{exp}(-\frac{t}{\tau_{2}})\,+\,A_{3}\textnormal{exp}(-\frac{t}{\tau_{3}})$

\begin{figure}[ht]
 \includegraphics[scale=0.6]{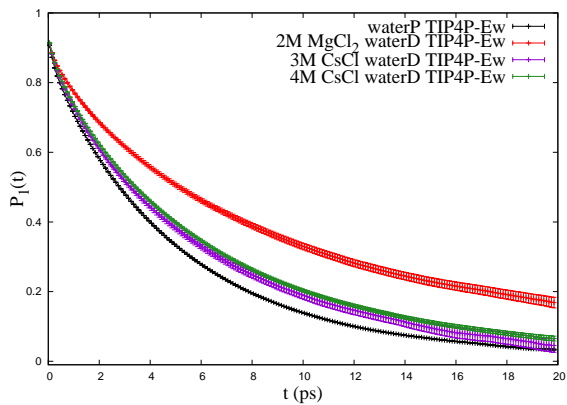}
 \caption{{\footnotesize Plots for P$_{\textnormal{1}}$(t) for \textit{waterD} comparing 3M and 4M CsCl with 2M MgCl$_2$. The black curve is for pure water using TIP4P-Ew model. 
 The error bars in all plots have been magnified 5 times}} 
\end{figure}

depending on the system. However, the attribution of specific physical processes to these time constants is not attempted in this work. 
A bi-exponential function ($A_3=0$) was the best 
fit for reorientational time curves for the pure water case.  In the presence of MgCl$_2$, a 
tri-exponential function was found to be more appropriate, with dynamics of defect water molecules 
introducing a new time scale into the problem. The decay times ($\tau_3$) are much larger than the 
slowest component of orientational relaxation for bulk water and show an increase with the concentration of 
MgCl$_2$ salt. These results are shown in TABLE 1. Simulation results show that for the largest 
concentration of MgCl$_2$ studied (4M), the longest mode in decay time for defect water molecules, which are 
beyond first hydration shell of any ions, is as high as 21 ps. The P$_{\textnormal{1}}$(t) curves for 
CsCl salt solutions are plotted in FIG. 3 and deviation from the case of pure water is much less 
significant compared to effects of MgCl$_2$ and independent of salt concentration. The 
P$_{\textnormal{1}}$(t) curves for CsCl are also fit with a tri-exponential function and the values are 
given in TABLE 1. The corresponding reorientational time constants obtained for P$_{\textnormal{2}}$(t) for sub-population \textit{waterD} for TIP4P-Ew and
P$_{\textnormal{1/2}}$(t) for TIP3P water models are given in \textit{Supp.~Info.} (TABLE. SII)~\cite{supplement}. The anion used in both the salts, Cl$^-$, is known to be weakly hydrated ion and 
such anions are expected to affect the dynamics of OH vector preferentially over dipole vectors of 
water molecule~\cite{tiel2}. Thus the observed difference in reorientational dynamics between the 
two cations studied in this work, Mg$^{2+}$ and Cs$^+$ reflect the difference between a strongly 
hydrated vs a weakly hydrated cation, while both have the same counterion. Similar differences 
between Mg$^{2+}$ and Cs$^+$ are observed for the sub-population of water molecules \textit{waterD2} as 
well and the results are included in \textit{Supp.~Info.} (FIG. SIV)~\cite{supplement} for both water models used. These results 
seem to be consistent with the ordering of cations in well-known Hofmeister series~\cite{tiel2}. 
\begin{table*}[ht] \footnotesize \begin{tabular}{ |c|c|c|c|c|c|c|c| }
 \hline
 salt & sub-population & $A_1$ & $\tau_1$ & $A_2$ & $\tau_2$ & $A_3$ & $\tau_3$  \\
 \hline
 none &  & 0.078($\pm$0.002)& 0.329($\pm$0.052) & 0.848($\pm$0.002) & 5.333($\pm$0.034) &  &  \\
 \hline
 \multirow{3}{*}{MgCl$_2$} & (2M,\textit{waterD}) & 0.052($\pm$0.004) & 0.209($\pm$0.058) & 0.262($\pm$0.003) & 3.538($\pm$0.031) & 0.615($\pm$0.002) & 15.027($\pm$0.002) \\
    & (3M,\textit{waterD}) & 0.049($\pm$0.003) & 0.192($\pm$0.028) & 0.222($\pm$0.003) & 3.010($\pm$0.016) & 0.656($\pm$0.003) & 17.654($\pm$0.001) \\
    & (4M,\textit{waterD}) & 0.050($\pm$0.002) & 0.206($\pm$0.024) & 0.197($\pm$0.002) & 2.980($\pm$0.017) & 0.682($\pm$0.002) & 20.818($\pm$0.001) \\
 \hline
 \multirow{2}{*}{CsCl} & (3M,\textit{waterD}) & 0.044($\pm$0.004) & 0.188($\pm$0.307) & 0.169($\pm$0.005) & 1.917($\pm$0.039) & 0.720($\pm$0.003) & 7.476($\pm$0.002) \\
    & (4M,\textit{waterD}) & 0.052($\pm$0.003) & 0.232($\pm$0.031) & 0.235($\pm$0.009) & 2.788($\pm$0.022) & 0.644($\pm$0.002) & 8.506($\pm$0.002) \\
 \hline
 \end{tabular}
\caption{\footnotesize Water reorientational time constants (for P$_{\textnormal{1}}$(t)) with TIP4P-Ew water model (under NVE conditions following NPT (T $=$ 298K) simulations) for pure water
molecules and \textit{waterD} sub-population in the presence of salt using tri-exponential fits. All time constants are in picoseconds.}
\end{table*}
\normalsize

Two dynamical quantities, which can be measured experimentally, related to ion hydration are the 
hydration number ($N_{\vec{p}}$), defined as the number of moles of slow water dipoles per mole of 
dissolved salt and fraction of slow water molecules relative to bulk-like water 
($f_{\frac{\textnormal{slow}}{\textnormal{bulk}}})$~\cite{tiel,tiel2}. The slow water molecules 
identified in the experiments are independent of structural definition of hydration shells and can 
contain water molecules within and outside the first hydration shells of ions. To be consistent 
with the experiments, a similar definition of slow water molecules was adopted which entails including water 
molecules in the first hydration shell of cations along with two sub-populations \textit{waterD} and \textit{waterD2}. 
Further, simulations allow a new dynamical quantity 
$f_{(\frac{\textnormal{slow}}{\textnormal{bulk}},\textnormal{defect})}$ to be computed, which is 
difficult to measure experimentally. 
$f_{(\frac{\textnormal{slow}}{\textnormal{bulk}},\textnormal{defect})}$ measures the fraction of 
slow water molecules beyond first hydration shell of ions relative to bulk-like water. The $N_{\vec{p}}$, 
$f_{\frac{\textnormal{slow}}{\textnormal{bulk}}}$ and 
$f_{(\frac{\textnormal{slow}}{\textnormal{bulk}},\textnormal{defect})}$ for both MgCl$_2$ and CsCl 
solutions are given in TABLE II. The corresponding numbers for TIP3P water model are given in \textit{Supp.~Info.} (TABLE SIII)~\cite{supplement}. It has been suggested that the typical hydration number, 
$N_{\vec{p}}$, for many ions is around 6 and any value greater than this number indicates presence 
of long-range effects of ions~\cite{tiel}. Experiments on salt solutions containing both strongly 
hydrated cations and anions show a large $N_{\vec{p}}$ value of the order of 18 and this 
has been suggested as a strong indication of cooperative slow down of water dynamics beyond first 
hydration shells of such ions~\cite{tiel}. In the present study the $N_{\vec{p}}$ values for MgCl$_2$ for all 
the three concentrations studied is 7, indicating the existence of the long-range effect of 
Mg$^{2+}$ ions in the presence of Cl$^-$, albeit weaker than when the counterion is SO$_4^{2-}$. The 
$N_{\vec{p}}$ values for CsCl were found to be less than 6.

From TABLE II, it can also be seen that the fraction of slow water molecules relative to bulk-like water, 
beyond first hydration shell 
($f_{(\frac{\textnormal{slow}}{\textnormal{bulk}},\textnormal{defect})}$) for 2M MgCl$_2$ is only 
0.06. This small value may suggest why long-range effects of strongly hydrated cations were not 
conclusively found in earlier simulations. A snap shot of MgCl$_2$-TIP4P-Ew water system for 2M salt 
concentration is shown in FIG. 4. The defect water molecules, which are beyond the first hydration shell of any 
ions are shown as van der Waals spheres and it can be seen that they form a small fraction of total 
number of water molecules in the system. Values of $f_{\frac{\textnormal{slow}}{\textnormal{bulk}}}$ 
as a function of radial distance from cations was computed and are given in TABLE SIV in 
\textit{Supp.~Info.}~\cite{supplement}. Within 2.5 $\mathring{A}$, a typical radial distance defining first 
hydration shell, the fraction of slow water molecules to bulk 
$f_{\frac{\textnormal{slow}}{\textnormal{bulk}}}$ for 2M Mg Cl$_2$ concentration is 0.89. This value 
drops to 0.22 just above 3 $\mathring{A}$ and progressively decreases beyond first hydration shell. 
However the non-zero values of $f_{\frac{\textnormal{slow}}{\textnormal{bulk}}}$ well beyond first 
and second hydrations shells of Mg$^{2+}$ ions even at 2M concentration indicate delocalised 
presence of slow water molecules. Similartrends are observed forTIP3P water model as well.
These results suggest that classifying water molecules around multiple ions, at moderate to high concentrations, using 
an oft-used definition of radially varying hydration shells may not be able to capture this 
small fraction of slow water molecules. The definition of defect water molecules \textit{waterD} used 
in this study can capture these small fraction of slow water molecules and suggest long-range 
effects of strongly hydrated cations such as Mg$^{2+}$. It can also be seen from the TABLE II that 
this fraction of defect water molecules increases with concentration and a significant jump occurs 
in such fraction from 2M to 4M (0.19 to 0.80). This suggests that a global network of defect water molecules 
may occur at higher salt concentrations. The values of 
$f_{\frac{\textnormal{slow}}{\textnormal{bulk}}}$ and 
$f_{(\frac{\textnormal{slow}}{\textnormal{bulk}},\textnormal{defect})}$ for CsCl salt solutions are 
very small, consistent with results in FIG. 3.

\begin{figure}[h]
 \includegraphics[scale=0.5]{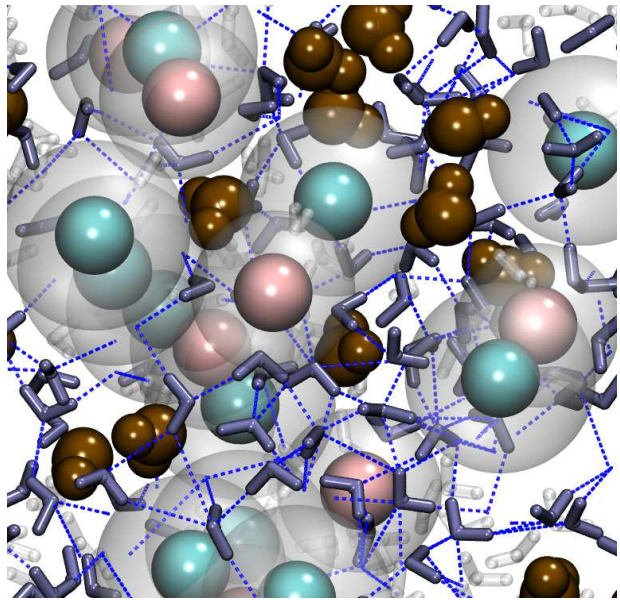}
 \caption{{\footnotesize Snapshot of part of MgCl$_2$-water system at 2M salt concentration. The Mg$^{2+}$ and Cl$^-$ ions are shown as cyan and pink spheres. 
The first hydration shells (2.5 $\mathring{A}$ radius) around the ions are also shown as transparent spheres. The defect water molecules, \textit{waterD}, which are not
within the first hydration shell are shown in vdW representation in brown and the other water molecules are shown in stick representation (transparent for
water molecules in the first solvation shell of ions). The hydrogen bond network among water molecules is also shown.}}
\end{figure}
\begin{table}[h]
\footnotesize
\begin{tabular}{ |c|c|c|c| }
 \hline
 salt,conc. & $N_{\vec{p}}$ & $f_{\frac{\textnormal{slow}}{\textnormal{bulk}}}$ & $f_{(\frac{\textnormal{slow}}{\textnormal{bulk}},\textnormal{defect})}$ \\
 \hline
 MgCl$_2$, 2M & 7.28 & 0.19 & 0.06 \\
 \hline
 MgCl$_2$, 3M & 7.05 & 0.41 & 0.26 \\
  \hline
 MgCl$_2$, 4M & 6.71 & 0.80 & 0.34 \\
  \hline
 CsCl, 3M & 5.91 & 0.05 & 0.03 \\
  \hline
 CsCl, 4M & 5.80 & 0.15 & 0.10 \\
\hline
 \end{tabular}
\caption{\footnotesize Approximate values of quantities $N_{\vec{p}}$, $f_{\frac{\textnormal{slow}}{\textnormal{bulk}}}$ 
and $f_{(\frac{\textnormal{slow}}{\textnormal{bulk}},\textnormal{defect})}$ obtained with TIP4P-Ew water model  (under NVE conditions following NPT (T $=$ 298K) simulations). 
The data are averaged over 5000 frames and 0.5 ns of simulation time.}
\end{table}
\normalsize

\section{Conclusion} To summarize, MD simulation results in this study support the 
concentration dependent effects of strongly hydrated ion species (Mg$^{2+}$) on the reorientational 
dynamics of water molecules beyond the first hydration shell. A likely mechanism for the same has 
been suggested in terms of salt induced defects in the underlying hydrogen bond network 
of water, which is in agreement with concepts of ion induced patterning of water at long distances. 
While the actual number of water molecules beyond first hydration shell of ions that exhibit slow 
reorientational times is a small fraction of the total water molecules, especially at 
low concentrations, they have been found at large spatial separations from the ions. The fractional number 
has been observed to 
increase monotonically with increase in the concentration of MgCl$_2$. It is to be noted that the 
long-range effect of Mg$^{2+}$ ions in the presence of a weakly hydrated counterion Cl$^-$ is 
smaller than experimentally observed effects in the presence of SO$_4^{2-}$~\cite{tiel}, but not 
insignificant. Comparison with weakly hydrated 
cationic species (Cs$^+$) has been seen to be in agreement with the Hofmeister series. The results in the present study
are obtained with two non-polarizable models of water and these results are 
expected to be enhanced when polarizable models of water are used and are a part of future studies.

\section{Acknowledgments}
All the simulations in this work have been carried out on 1024-cpu Annapurna cluster at The Institute of Mathematical Sciences, 
Chennai, India.
 
\end{document}